# Fast Maximum-Likelihood Decoding of the Golden Code

Mohanned O. Sinnokrot and John R. Barry

School of ECE, Georgia Institute of Technology, Atlanta, GA 30332 USA, barry@ece.gatech.edu

*Abstract* — **The golden code is a full-rate full-diversity space-time code for two transmit antennas that has a maximal coding gain. Because each codeword conveys four information symbols from an *M*-ary quadrature-amplitude modulation alphabet, the complexity of an exhaustive search decoder is proportional to $M^4$. In this paper we present a new fast algorithm for maximum-likelihood decoding of the golden code that has a worst-case complexity of only $\mathcal{O}(2M^{2.5})$. We also present an efficient implementation of the fast decoder that exhibits a low average complexity. Finally, in contrast to the overlaid Alamouti codes, which lose their fast decodability property on time-varying channels, we show that the golden code is fast decodable on both quasistatic and rapid time-varying channels.**

I. INTRODUCTION

A WIRELESS communication system having multiple antennas at both the transmitter and receiver is capable of achieving higher data rates and greater robustness to fading than a single-antenna system. These gains can be achieved in part through the use of space-time coding [1,2]. The golden code is a space-time code for two transmit and two or more receive antennas that was proposed independently in [3] and [4]. The golden code has many advantages: it is a full-rate code (since it transmits four complex information symbols in two signaling intervals); it is a full-diversity code; it has a maximal coding gain; and in terms of the SNR required to achieve a target error probability, it performs better than all previously reported full-rate codes with two transmit antennas. Furthermore, the coding gain of the golden code is independent of the alphabet size, which has two benefits: first, it ensures that the golden code achieves the full diversity-multiplexing frontier of Zheng and Tse [5] (see also [6]); and second, from a practical standpoint it makes the golden code compatible with adaptive modulation. For these reasons, the golden code has been incorporated into the 802.16e WiMAX standard [7].

The golden code comes in three variations: the Belfiore-Rekaya-Viterbo golden code [3], the Dayal-Varanasi golden code [4], and the WiMAX golden code [7]. These variations are isomorphic, in the sense that one can be transformed into another by multiplying on the left and right by unitary matrices [4]. Since the determinant is invariant to such transformations, all three variations have identical rate, diversity, and coding gain. Furthermore, we will show that they all have the same decoding complexity. For the sake of concreteness, this paper focuses on decoding of the Dayal-Varanasi golden code [4]. Nevertheless, we emphasize that the results presented here are applicable all three variations [3][4][7].

The golden code applied to a system with two receive antennas leads to an *effective* four-input four-output MIMO channel that maps each block of four $M$-ary information symbols to a corresponding vector of four complex-valued received samples [8]. An exhaustive search maximum-likelihood (ML) decoder would consider each of the $M^4$ possible input vectors in turn and choose as its decision the one that best represents the channel output in a minimum-Euclidean-distance sense. Therefore, the complexity of such an ML decoder is proportional to $M^4$. Significant reductions in *average* complexity are possible by adopting a tree-based ML decoder such as a sphere decoder. However, for certain alphabets, the *worst-case* complexity of a sphere decoder and an arbitrary channel is no better than that of an exhaustive search. For this reason, it has been widely reported that the worst-case decoding complexity of the golden code grows with the fourth power of the signal constellation size [9,10,11,12,13].

The worst-case complexity depends dramatically on the alphabet type. In particular, when an arbitrary $M$-ary alphabet is replaced by the practically important special case of a QAM alphabet, the worst-case complexity of a tree-based sphere decoder for an arbitrary four-input MIMO channel drops from $\mathcal{O}(M^4)$ to $\mathcal{O}(M^3)$. To see why, consider the problem of finding the "best" leaf node stemming from a particular node at the third level of the four-level tree. While one could in principle exhaustively search all $M$ possibilities in turn, the problem can be solved efficiently using a single instance of a decision device, or *slicer*. And for certain alphabets, like QAM, the complexity of a slicer does not grow with the size of the alphabet; rather, the worst-case complexity of a QAM slicer is $\mathcal{O}(1)$. Specifically, a QAM slicer can be implemented as a pair of PAM slicers, with each requiring a single multiply, a single rounding operation, a single addition, and a single hard-limiting operation, none of which depends on $M$.

The perception that maximum-likelihood decoding of the golden code has high complexity has had two effects: First, it has motivated a search for suboptimal decoders for the golden code with reduced complexity and near-ML performance [14][15][16][17]. Second, it has motivated a search for alternatives to the golden code with similar performance but whose maximum-likelihood decoder has lower complexity [9][10][18]. In particular, two families of space-time codes based on the Alamouti code were proposed in [9][10] and [18]; we will refer to these codes as *overlaid Alamouti* codes, because these rate-two codes can be viewed as the sum of a conventional rate-one Alamouti code with a second modified rate-one Alamouti code. The overlaid Alamouti codes share the same full rate, full diversity, and nonvanishing determinant properties of the golden code, albeit with a small coding gain penalty of a fraction of a dB. Moreover, on quasistatic channels, the overlaid Alamouti codes of [9][10] and [18] admit a maximum-likelihood decoder whose worst-case complexity is only $\mathcal{O}(M^2)$ for $M$-ary QAM alphabets.

In this paper, we describe a new maximum-likelihood decoding algorithm for the golden code with $M$-ary QAM whose worst-case complexity is $\mathcal{O}(2M^{2.5})$. We also present an efficient implementation that has low average complexity. The proposed decoder exploits a key property of the effective channel induced by the golden code, namely that the inner product between the first and second columns is real. Furthermore, we will show that the golden code retains its fast-decodability property on time-varying channels. In contrast, we will see that the overlaid Alamouti codes of [9][10] and [18] lose their fast-decoding property on time-varying channels. Reduced complexity decoding for time-varying channels is particularly important in mobile applications, where user mobility can lead to rapid channel variations.

The remainder of the paper is organized as follows. In Section II, we review the golden code. In Section III, we introduce a new fast maximum-likelihood decoder for the golden code and show that the fast decodability is retained on time-varying channels. In Section IV, we show that the overlaid Alamouti codes are not fast decodable on time-varying channels. In Section V, we compare the average complexity of the proposed detector to a conventional golden code detector. We conclude the paper in Section VI.

## II. THE GOLDEN CODE

The golden code transmits four complex information symbols $\{x_1, x_2, x_3, x_4\}$ over two symbol periods from two antennas, so that the rate of the space-time code is two symbols per signaling interval. The transmitted codeword can be expressed as a $2 \times 2$ matrix:

$$\mathbf{C} = \begin{bmatrix} c_1[1] & c_2[1] \\ c_1[2] & c_2[2] \end{bmatrix}, \tag{1}$$

where $c_i[k]$ denotes the symbol transmitted from antenna $i \in \{1, 2\}$ at time $k \in \{1, 2\}$. In particular, the Dayal-Varanasi golden code encodes one pair of information symbols $\mathbf{a} = [x_1, x_2]^T$ onto the main diagonal of $\mathbf{C}$, and it encodes a second pair of information symbols $\mathbf{b} = [x_3, x_4]^T$ onto the off-diagonal. Specifically, the Dayal-Varanasi golden code is [4]:

$$\mathbf{C} = \begin{bmatrix} \tilde{a}_1 & 0 \\ 0 & \tilde{a}_2 \end{bmatrix} + \varphi \begin{bmatrix} 0 & \tilde{b}_1 \\ \tilde{b}_2 & 0 \end{bmatrix} \tag{2}$$

where:

$$\tilde{\mathbf{a}} = \mathbf{M}\mathbf{a}, \quad \tilde{\mathbf{b}} = \mathbf{M}\mathbf{b}, \quad \mathbf{M} = \begin{bmatrix} \cos(\theta) & \sin(\theta) \\ -\sin(\theta) & \cos(\theta) \end{bmatrix}, \quad \theta = \frac{1}{2}\tan^{-1}(2), \quad \text{and} \quad \varphi = e^{j\pi/4}. \tag{3}$$

As can be seen from (2), the golden code can be viewed as the sum of two rate-one diagonal algebraic space-time (DAST) codes [19].

Our model for the received signal $y_j[k]$ at receive antenna $j$ at time $k$ is given by:

$$y_j[k] = \sum_{i=1}^{2} c_i[k] h_{i,j}[k] + n_j[k], \qquad (4)$$

where $n_j[k]$ is the complex additive-white Gaussian noise at receive antenna $j$ at time $k$, and $h_{i,j}[k]$ is the channel coefficient between the $i$-th transmit antenna and $j$-th receive antenna at time $k$. For quasistatic fading, $h_{i,j}[k] = h_{i,j}$ is independent of time $k$.

## III. A Fast ML Decoder for the Golden Code

We begin by describing the effective channel matrix induced by the golden code. We then establish a key property of this matrix, and describe a maximum-likelihood decoder that exploits the key property to reduce complexity.

### A. The Effective Channel Matrix and its Key Property

Substituting the definition of the golden code from (2) and (3) into (4), the vector of samples $\mathbf{y} = [y_1[1], y_1[2], y_2[1], y_2[2]]^T$ received at a receiver with two antennas at the two time instances can be written as the output of an effective four-input four-output channel:

$$\mathbf{y} = \mathbf{H}\mathbf{x} + \mathbf{n}, \qquad (5)$$

where $\mathbf{x} = [x_1, \ldots x_4]^T$ is the vector of information symbols, $\mathbf{n} = [n_1[1], \ldots n_2[2]]^T$ is the noise, and where $\mathbf{H} = \overline{\mathbf{H}} \Psi$ is the *effective channel matrix*:

$$\mathbf{H} = \underbrace{\begin{bmatrix} h_{11}[1] & 0 & \varphi h_{21}[1] & 0 \\ 0 & h_{21}[2] & 0 & \varphi h_{11}[2] \\ h_{12}[1] & 0 & \varphi h_{22}[1] & 0 \\ 0 & h_{22}[2] & 0 & \varphi h_{12}[2] \end{bmatrix}}_{\overline{\mathbf{H}}} \underbrace{\begin{bmatrix} c & s & 0 & 0 \\ -s & c & 0 & 0 \\ 0 & 0 & c & s \\ 0 & 0 & -s & c \end{bmatrix}}_{\Psi}, \qquad (6)$$

where $c = \cos(\theta)$, $s = \sin(\theta)$, $\varphi = e^{j\pi/4}$, and $\theta = \frac{1}{2}\tan^{-1}(2)$.

The structure of the golden code induces special properties in this effective matrix that we exploit to reduce decoding complexity. The following theorem relates these special properties to the orthogonal-triangular (QR) decomposition $\mathbf{H} = \mathbf{QR}$, which results from an application of the Gram-Schmidt procedure to the columns of $\mathbf{H} = [\mathbf{h}_1, \ldots \mathbf{h}_4]$, where $\mathbf{Q} = [\mathbf{q}_1, \ldots \mathbf{q}_4]$ is unitary and $\mathbf{R}$ is upper triangular with nonnegative real diagonal elements, so that the entry of $\mathbf{R}$ in row $i$ and column $j$ is $r_{i,j} = \mathbf{q}_i^* \mathbf{h}_j$.

**Theorem 1.** *(The Key Property)*: The $\mathbf{R}$ matrix in a QR decomposition $\mathbf{H} = \mathbf{QR}$ of the effective channel (6) has the form

$$\mathbf{R} = \begin{bmatrix} \mathbf{A} & \mathbf{B} \\ \mathbf{O} & \mathbf{D} \end{bmatrix}, \tag{7}$$

*where both of the upper triangular matrices $\mathbf{A}$ and $\mathbf{D}$ are entirely real.*

*Proof:* See Appendix A.

A few remarks:

- By construction, $\mathbf{A} = \begin{bmatrix} r_{11} & r_{12} \\ 0 & r_{22} \end{bmatrix}$ and $\mathbf{D} = \begin{bmatrix} r_{33} & r_{34} \\ 0 & r_{44} \end{bmatrix}$ are triangular with real diagonal entries, so the key property is essentially the fact that both $r_{1,2}$ and $r_{3,4}$ are real.

- To demonstrate that $r_{1,2} = \mathbf{h}_1^* \mathbf{h}_2 / \|\mathbf{h}_1\|$ is real, it is sufficient to show that the inner product between the first two columns of the effective channel matrix is real, a fact which is easily verified by direct computation:

$$\begin{aligned}\mathbf{h}_1^* \mathbf{h}_2 &= \cos(\theta)\sin(\theta)(|h_{11}[1]|^2 - |h_{21}[2]|^2 + |h_{12}[1]|^2 - |h_{22}[2]|^2) \\ &= \frac{1}{\sqrt{5}}(|h_{11}[1]|^2 + |h_{12}[1]|^2 - |h_{21}[2]|^2 - |h_{22}[2]|^2).\end{aligned} \tag{8}$$

- The theorem applies regardless of whether the channel is quasistatic or time-varying.

- The submatrix $\mathbf{B}$ is not mentioned because all four of its entries are generally complex, regardless of whether the channel is quasistatic or time-varying.

- The fact that $r_{1,2}$ is real enables the decoder in the next section to reduce both the worst-case decoding complexity and the average decoding complexity. In contrast, the fact that $r_{3,4}$ is real enables only a reduction in average complexity. It has no impact on the worst-case complexity.

## B. A Fast ML Decoder for the Golden Code

If we define $\mathbf{z}_{12} = (z_1, z_2)^T$ and $\mathbf{z}_{34} = (z_3, z_4)^T$, where $\mathbf{z} = \mathbf{Q}^*\mathbf{y}$, then the maximum-likelihood decision minimizes the cost function

$$\begin{aligned} P(\mathbf{x}) &= \|\mathbf{y} - \mathbf{H}\mathbf{x}\|^2 \\ &= \|\mathbf{z} - \mathbf{R}\mathbf{x}\|^2 \\ &= \|\mathbf{z}_{12} - \mathbf{A}\mathbf{a} - \mathbf{B}\mathbf{b}\|^2 + \|\mathbf{z}_{34} - \mathbf{D}\mathbf{b}\|^2. \end{aligned} \quad (9)$$

The last equality follows from (7). Therefore, the ML decisions $\hat{\mathbf{a}}$ and $\hat{\mathbf{b}}$ can be found recursively using:

$$\hat{\mathbf{b}} = \mathrm{argmin}_{\mathbf{b} \in \mathcal{A}^2} \{\|\mathbf{z}_{12} - \mathbf{A}\mathbf{a}_*(\mathbf{b}) - \mathbf{B}\mathbf{b}\|^2 + \|\mathbf{z}_{34} - \mathbf{D}\mathbf{b}\|^2\}, \quad (10)$$

$$\hat{\mathbf{a}} = \mathbf{a}_*(\hat{\mathbf{b}}), \quad (11)$$

where

$$\mathbf{a}_*(\mathbf{b}) = \mathrm{argmin}_{\mathbf{a} \in \mathcal{A}^2} \|\mathbf{z}_{12} - \mathbf{A}\mathbf{a} - \mathbf{B}\mathbf{b}\|^2. \quad (12)$$

The function $\mathbf{a}_*(\mathbf{b})$ in (12) can be viewed as producing the best $\mathbf{a}$ for a given $\mathbf{b}$. With this interpretation, the optimization in (10) can be viewed as that of finding the best $\mathbf{b}$ when $\mathbf{a}$ is optimized.

The optimization in (12) is equivalent to ML detection for a channel $\mathbf{A}$ with two QAM inputs and an output of:

$$\mathbf{v} = \mathbf{z}_{12} - \mathbf{B}\mathbf{b}. \quad (13)$$

It can thus be solved by a sphere detector applied to a two-level tree. As discussed in the introduction, with two QAM inputs and without any constraints on $\mathbf{A}$, its worst-case complexity would be $\mathcal{O}(M)$. But the golden code induces the special property that $\mathbf{A}$ is entirely real, which enables (12) to determine the real components of $\mathbf{a}$ *independently* from its imaginary components. Specifically, the fact that $\mathbf{A}$ is real enables us to rewrite (12) as:

$$\mathbf{a}_*(\mathbf{b}) = \mathrm{argmin}_{\mathbf{a} \in \mathcal{A}^2} \{\|\mathbf{v}^R - \mathbf{A}\mathbf{a}^R\|^2 + \|\mathbf{v}^I - \mathbf{A}\mathbf{a}^I\|^2\} \quad (14)$$

$$= \mathrm{argmin}_{\mathbf{a}^R \in (\mathcal{A}^R)^2} \{\|\mathbf{v}^R - \mathbf{A}\mathbf{a}^R\|^2\} + j \cdot \mathrm{argmin}_{\mathbf{a}^I \in (\mathcal{A}^I)^2} \{\|\mathbf{v}^I - \mathbf{A}\mathbf{a}^I\|^2\}. \quad (15)$$

(Throughout the paper we use superscripts $R$ and $I$ denote the real and imaginary components, respectively, so that $\mathbf{v}^R = \text{Re}\{\mathbf{v}\}$ and $\mathbf{a}^I = \text{Im}\{\mathbf{a}\}$). Thus, the optimization in (12) decomposes into the two parallel optimizations of (15). Since each optimization in (15) is equivalent to ML detection for a *real* channel with two $\sqrt{M}$-PAM inputs, each has a worst-case complexity of $\mathcal{O}(\sqrt{M})$. Thus, the overall complexity of (15) is $\mathcal{O}(2\sqrt{M})$.

We have thus shown how an ML decoder can be implemented for the golden code with a worst-case complexity of $\mathcal{O}(2M^{2.5})$: as described in (10), the ML decision can be found by stepping through each of the $M^2$ candidate values for $\mathbf{b}$, and for each implement the $\mathcal{O}(2\sqrt{M})$ optimization of (15). A more efficient implementation is described in the next section.

*C. A Faster ML Decoder with Low Average Complexity*

The ML decoder of the previous section is fast, in the sense that it has a worst-case complexity of $\mathcal{O}(2M^{2.5})$, but its average complexity is unnecessarily high. In this section we describe an efficient implementation of an ML decoder for the golden code that has both low average complexity and a worst-case complexity of $\mathcal{O}(2M^{2.5})$.

A conventional sphere decoder for the golden code is based on a four-level tree, with a different $x_i$ associated with each level. In contrast, as illustrated in Fig. 1, we propose a four-level tree that associates $\mathbf{b}^R = (x_3^R, x_4^R)$ with the first level, $\mathbf{b}^I = (x_3^I, x_4^I)$ with the second level, $\mathbf{a}^R = (x_1^R, x_2^R)$ with the third level, and $\mathbf{a}^I = (x_1^I, x_2^I)$ with the fourth level. This new tree is a direct result of the fact that $\mathbf{A}$ and $\mathbf{D}$ are real (Theorem 1), which allows us to rewrite the ML cost function from (9) as

$$P(\mathbf{x}) = \underbrace{\|\mathbf{v}^I - \mathbf{A}\mathbf{a}^I\|^2}_{P_1} + \underbrace{\|\mathbf{v}^R - \mathbf{A}\mathbf{a}^R\|^2}_{P_2} + \underbrace{\|\mathbf{z}_{34}^I - \mathbf{D}\mathbf{b}^I\|^2}_{P_3} + \underbrace{\|\mathbf{z}_{34}^R - \mathbf{D}\mathbf{b}^R\|^2}_{P_4}. \quad (16)$$

Thus, as illustrated in Fig. 1, (16) shows that the total cost of a leaf node $\mathbf{x}$ decomposes into the sum of four branch metrics, where $P_i$ denotes the branch metric for a branch at the $(4-i)$-th stage of the tree.

Besides inducing a new tree structure, the fact that **D** is real also leads to a significant reduction in the complexity of Schnorr-Euchner sorting for the first two stages of the tree. Specifically, the fact that **D** is real leads to a second-stage branch metric $P_3$ that is *independent* of the starting node ($\mathbf{b}^R$). Therefore, we can perform a single sort for the symbol pair ($\mathbf{b}^R$) emanating from the root, and *simultaneously* a single sort for the symbol pair ($\mathbf{b}^I$) emanating from its children. In contrast, if **D** were complex, we would require one sorting operation for the root node, and then a separate sorting operation for each visited child. In the worst case, when every child is visited, there would be a total of $M + 1$ sorting operations required for the first two tree stages. Thus, the fact that **D** is real enables us to reduce the number of sorts from as many as $M + 1$ to only two. The computational savings is significant, especially for large alphabets.

The pseudocode of an efficient implementation of the proposed maximum-likelihood golden code detector is shown in Fig. 2. The first five lines represent initializations. In particular, the first two lines are a QR decomposition of the effective channel matrix in (6) and the computation of **z** in (9). The squared sphere radius $\hat{P}$, which represents the smallest cost (16) encountered so far, is initialized to infinity to ensure ML decoding (line 3). Sorting or Schnorr-Euchner enumeration is used for faster convergence. Only two sorting operations (line 4 and line 5) are required. In the pseudocode, the complex QAM alphabet $\mathcal{A}$ is represented by an ordered list, so that $\mathcal{A}(k)$ indexes the $k$-th symbol in the list.

We next describe the remainder of algorithm, which can be interpreted as a two-level *complex* sphere decoder to choose the symbol pair $\mathbf{b} = (x_3, x_4)^T$, followed by an independent pair of two-level *real* sphere decoders that separately decode $\mathbf{a}^R = (x_1^R, x_2^R)^T$ and $\mathbf{a}^I = (x_1^I, x_2^I)^T$.

The two-level complex sphere decoder incorporates two common optimizations: radius update (line 33) and pruning (line 7, line 9). While these optimizations do not affect the worst-case complexity, they affect the average complexity significantly. The first level of the complex sphere decoder considers candidate pairs $\mathbf{b}^R$ in ascending order of their branch metric $P_4$ (line 6). The second level of the complex sphere decoder considers candidate pairs $\mathbf{b}^I$ in ascending order of their branch metric $P_3$ (line 8). After forming **b** (line 10), the decoder removes the interference caused by **b** and forms the two intermediate variables $v_1$ and $v_2$ of (13), which are functions of the symbols $x_1$ and $x_2$ only (line 11 and line 12). Following the two-level complex sphere decoder and interference cancellation, the decoder decides on the symbol pairs $\mathbf{a}^R$ and $\mathbf{a}^I$ separately using an independent pair of two-level real sphere decoders.

The function `list` is used to implement Schnorr-Euchner sorting for the final two stages of the tree; it returns a list of candidate symbols drawn from the $\sqrt{M}$-ary PAM alphabet $\mathcal{A}^R$, sorted in ascending order of distance to the input argument. As described in [20], it can be implemented efficiently using a table lookup, requiring only a single rounding operation and a single addition.

After initializing the sphere radius for decoding $\mathbf{a}^R = (x_1^R, x_2^R)^T$ (line 13) and forming the sorted list of best candidate symbols (line 14), the real sphere decoder chooses the symbol $x_2^R$ that has the lowest branch metric $P_2$ (line 16). The interference from the symbol $x_2^R$ is then subtracted (line 18) and a decision is made on the symbol $x_1^R$ using the PAM slicer $Q(\,\cdot\,)$ (line 19); the slicer function $Q(x)$ returns the symbol from the PAM alphabet $\mathcal{A}^R$ that is closest to $x$. The branch-metric sum $P_2$ for the current candidate symbol pair $\mathbf{a}^R$ is computed in line 20, and radius update occurs if it is less than the previous smallest value $\hat{P}_2$ (line 21). Similar to the complex sphere decoder, the real sphere decoder includes pruning and sphere radius update (line 17 and line 21).

Decoding the symbol pair $\mathbf{a}^I$ follows identically to the decoding of the symbol pair $\mathbf{a}^R$ and is shown in line 24 through line 31. We emphasize that $\mathbf{a}^I$ is decoded separately from $\mathbf{a}^R$. The overall cost $P$ for the current candidate symbol vector is updated in line 32. Radius update and best candidate vector update occurs if the current cost $P$ is less than the previous smallest cost $\hat{P}$ (line 33).

The algorithm could be embellished to further reduce average complexity. For example, instead of sorting the entire alphabet as in line 4 and line 5, a sort-as-you-go approach would be more efficient. Furthermore, for faster convergence of the search algorithm, the QR decomposition in line 1 could permute the columns of **H**. For the sake of clarity of exposition, however, we have chosen not to include such refinements in Fig. 2. Such refinements, which have no effect on the worst-case complexity, are well-known in the literature and their application to the pseudocode of Fig. 2 is straightforward.

Regarding the possibility of permuting the columns of **H** for speeding convergence, we note that fast decoding is possible only for following permutations {1, 2, 3, 4}, {1, 2, 4, 3}, {2, 1, 3, 4}, {2, 1, 4, 3}, {3, 4, 1, 2}, {3, 4, 2, 1}, {4, 3, 1, 2} and {4, 3, 2, 1}. These are the only permutations for which the key property of Theorem 1 will still hold. The same restriction on the allowable permutations for reduced complexity decoding also applies to the overlaid Alamouti codes.

We remark that a quasistatic channel does not offer any additional reduction in decoding complexity, as compared to a time-varying channel. This is a direct result of the fact that the entries of **B** in (7) are generally complex, regardless of whether the channel is quasistatic or time-varying.

### D. Golden Code Variations

In this section we argue that the proposed fast ML decoder, although presented in the context of the Dayal-Varanasi version of the golden code [4], is equally applicable to the Belfiore-Rekaya-Viterbo [3] and WiMAX [7] versions of the golden code.

Substituting the definition of the Belfiore-Rekaya-Viterbo golden code from [3, eqn. (9)] into (4), the vector of samples received at a receiver with two antennas will again be given by (5), but with a new effective channel matrix of the form:

$$\mathbf{H} = \begin{bmatrix} h_{11}[1] & 0 & h_{21}[1] & 0 \\ 0 & h_{21}[2] & 0 & ih_{11}[2] \\ h_{12}[1] & 0 & h_{22}[1] & 0 \\ 0 & h_{22}[2] & 0 & ih_{12}[2] \end{bmatrix} \begin{bmatrix} c-si & 0 & 0 & 0 \\ 0 & s+ci & 0 & 0 \\ 0 & 0 & c-si & 0 \\ 0 & 0 & 0 & s+ci \end{bmatrix} \begin{bmatrix} c & s & 0 & 0 \\ -s & c & 0 & 0 \\ 0 & 0 & c & s \\ 0 & 0 & -s & c \end{bmatrix}, \quad (17)$$

where $c = \cos(\theta)$, $s = \sin(\theta)$, and $\theta = \frac{1}{2}\tan^{-1}(2)$. The information symbols $[b, a, d, c]$ in [3] have been relabeled as $[x_1, x_2, x_3, x_4]$ in (5). Since $c - si$ and $s + ci$ have unity magnitude, we can transform this effective matrix into the one in (6) simply by rotating the channel coefficients $h_{i,j}[k]$. These rotations have no impact on complexity. In particular, the real coefficients in the **R** matrix will remain real, even after rotation. Therefore, the proposed fast ML decoder is applicable.

Similarly, substituting the definition of the WiMAX golden code from [7, Sect. 8.4.8.3.3] (matrix C) into (4) will again yield (5), but with a new effective channel matrix of the form:

$$\mathbf{H} = \begin{bmatrix} h_{11}[1] & 0 & h_{21}[1] & 0 \\ 0 & -ih_{21}[2] & 0 & -h_{11}[2] \\ h_{12}[1] & 0 & h_{22}[1] & 0 \\ 0 & -ih_{22}[2] & 0 & -h_{12}[2] \end{bmatrix} \begin{bmatrix} c & s & 0 & 0 \\ -s & c & 0 & 0 \\ 0 & 0 & c & s \\ 0 & 0 & -s & c \end{bmatrix}. \qquad (18)$$

The information symbol vector $[S_i, iS_{i+3}, S_{i+1}, -S_{i+2}]$ in [7] has been relabeled to $[x_1, x_2, x_3, x_4]$ in (5). Just as before, this effective matrix differs from that in (6) only by the rotated channel coefficients. Therefore, the proposed fast ML decoder is again applicable to this version of the golden code.

### E. Extension to Other Alphabets using Coordinate Interleaving

The proposed fast decoder exploits the property of QAM that its real and imaginary components may be decoded separately. When the real and imaginary components are not separately decodable, as is the case for PSK, hexagonal, and cross-QAM alphabets, the real symbol pair $\mathbf{a}^R$ cannot be decoded separately from the imaginary symbol pair $\mathbf{a}^I$. This leads to a worst-case decoding complexity of $\mathcal{O}(M^3)$, the same as a conventional ML decoder.

The reduced decoding complexity of the golden code can be extended to arbitrary alphabets by interleaving the coordinates prior to encoding [21]. This so-called *interleaved* golden code allows the receiver to separate the decoding of the real and imaginary components of $x_1$ from the real and imaginary components of $x_2$. In other words, after cancelling the interference from symbols $x_3$ and $x_4$, the symbol $x_1$ is separately decodable from $x_2$. This key property leads to a worst-case decoding complexity of $\mathcal{O}(2M^{2.5})$, regardless of the alphabet type, and regardless of whether the channel is time varying. The interleaved golden code maintains the same coding gain and good performance of the golden code while extending its fast decodability from QAM to arbitrary alphabets.

## IV. ML Decoding of The Overlaid Alamouti Code

In this section we show that the overlaid Alamouti code of [18] loses its fast decodability when the channel varies with time. The same results hold for the overlaid Alamouti code of [10]; see [12] and the references therein.

The overlaid Alamouti space-time code of [18] is:

$$\mathbf{C} = \frac{1}{\sqrt{2}} \begin{bmatrix} x_1 & x_2 \\ -x_2^* & x_1^* \end{bmatrix} + \frac{1}{\sqrt{2}} \begin{bmatrix} 1 & 0 \\ 0 & -1 \end{bmatrix} \begin{bmatrix} u_1 & u_2 \\ -u_2^* & u_1^* \end{bmatrix}, \tag{19}$$

where

$$\begin{bmatrix} u_1 \\ u_2 \end{bmatrix} = \begin{bmatrix} \varphi_1 & \varphi_2 \\ -\varphi_2^* & \varphi_1^* \end{bmatrix} \begin{bmatrix} x_3 \\ x_4 \end{bmatrix}, \varphi_1 = \frac{1}{\sqrt{7}}(1+j), \text{ and } \varphi_2 = \frac{1}{\sqrt{7}}(1+2j). \tag{20}$$

Substituting the definition of the overlaid Alamouti code of [18] from (19) and (20) into (4), the vector of samples received at a receiver with two antennas at the two time instances can be written as:

$$\mathbf{y} = \begin{bmatrix} y_1[1] \\ y_1^*[2] \\ y_2[1] \\ y_2^*[2] \end{bmatrix} = \begin{bmatrix} h_{11}[1] & h_{21}[1] & \varphi_1 h_{11}[1] - \varphi_2^* h_{21}[1] & \varphi_2 h_{11}[1] + \varphi_1^* h_{21}[1] \\ h_{21}^*[2] & -h_{11}^*[2] & -\varphi_2^* h_{11}^*[2] + \varphi_1 h_{21}^*[2] & \varphi_1^* h_{11}^*[2] - \varphi_2 h_{21}^*[2] \\ h_{12}[1] & h_{22}[1] & \varphi_1 h_{12}[1] - \varphi_2^* h_{22}[1] & \varphi_2 h_{12}[1] + \varphi_1^* h_{22}[1] \\ h_{22}^*[2] & -h_{12}^*[2] & -\varphi_2^* h_{12}^*[2] + \varphi_1 h_{22}^*[2] & \varphi_1^* h_{12}^*[2] - \varphi_2 h_{22}^*[2] \end{bmatrix} \begin{bmatrix} x_1 \\ x_2 \\ x_3 \\ x_4 \end{bmatrix} + \begin{bmatrix} n_1[1] \\ n_1^*[2] \\ n_2[1] \\ n_2^*[2] \end{bmatrix}$$

$$= \mathbf{H}\mathbf{x} + \mathbf{n}. \tag{21}$$

From (21) we can see that a quasistatic channel causes the first column of $\mathbf{H}$ to be orthogonal to the second, and the third column to be orthogonal to the fourth. The implication on a QR decomposition $\mathbf{H} = \mathbf{Q}\mathbf{R}$ is that $r_{1,2} = r_{3,4} = 0$. The receiver can exploit the fact that $r_{1,2} = 0$ to reduce the worst-case decoding complexity as follows: the receiver chooses a pair $(x_3, x_4)$ and subtracts their contribution. Then, for every such pair, the receivers decides on the remaining symbols $x_1$ and $x_2$ separately. The separate decoding of $x_1$ and $x_2$ is possible because $r_{1,2} = 0$. For a square $M$-ary QAM alphabet, the worst-case decoding complexity is $\mathcal{O}(4M^2)$, where $M^2$ comes from the fact that there are $M^2$ ways to choose the symbol pair $(x_3, x_4)$ and $4\mathcal{O}(1)$ comes from the fact that the real and imaginary components of $x_1$ and $x_2$ are separately decodable and each has a decoding complexity of $\mathcal{O}(1)$.

Under time-varying fading, however, the orthogonality of the columns of **H** is lost, and the **R** matrix does not have any zero upper entries. In fact, all of the entries above the diagonal are complex in general. Therefore, ML detection has a worst-case decoding complexity of $\mathcal{O}(M^3)$. This result is not surprising when one considers that the Alamouti code cannot be decoded in a low-complexity manner when the channel varies with time, and the overlaid Alamouti codes are the sum of two Alamouti block codes.

V. NUMERICAL RESULTS

In Fig. 3 we compare the average complexity of the proposed fast ML decoder to a conventional ML decoder. The channel was modeled using (4) with quasistatic i.i.d. Rayleigh fading, with constant channel coefficients within each codeword block, and independent complex normal coefficients from block to block. The alphabet was 64-QAM. The fast ML decoder was implemented following the pseudocode of Fig. 2. The conventional ML decoder was implemented using a four-level complex sphere decoder with Schnorr-Euchner enumeration for fast convergence. Results are shown for two cases of channel matrix column ordering: no ordering and BLAST reordering.

As can be seen from Fig. 3, with no column ordering, the average complexity of the proposed fast ML decoder is about 45% less complex than a conventional ML decoder. With BLAST ordering, the proposed ML decoder is about 30% less complex than a conventional decoder.

The average complexity is quantified by the average number of nodes visited while searching the tree; this measure of complexity has the advantages of being simple and reasonably insensitive to the implementation details of the algorithm. Other complexity measures such as floating point operations (FLOPs) are quite sensitive to the implementation of the algorithm and can significantly exaggerate the performance of one algorithm compared to another. For example, the use of a look-up table for symbol sorting would significantly reduce the FLOP count, but it would not affect the convergence time or the average node count.

A significant drawback of the average node count is that it does not capture the complexity of the column ordering and Schnorr-Euchner symbol sorting. Therefore, beyond the advantages shown in Fig. 3, the proposed algorithm has three additional advantages that are not reflected in Fig. 3, namely:

- the proposed algorithm reduces the number of Schnorr-Euchner sort operations for the first two stages to only two, compared with a conventional decoder that can require as many as $M + 1$. The resulting complexity reduction can be significant, since sorting is an expensive operation.
- the proposed algorithm can avoid the high complexity of BLAST ordering without a high performance penalty.
- decoding of the symbol pairs $\mathbf{a}^R$ and $\mathbf{a}^I$ can be done in parallel, reducing decoding latency.

## VI. CONCLUSIONS

We have proposed a maximum-likelihood decoding algorithm for the golden code and its variants with $M$-ary QAM whose worst-case decoding complexity is $\mathcal{O}(2M^{2.5})$. For large alphabets, this represents a significant reduction in complexity compared to the worst-case of $\mathcal{O}(M^3)$ for a conventional detector. We have presented an efficient implementation that was shown to significantly outperform a conventional detector in terms of average complexity. Finally, unlike the alternatives to the golden code based on overlaying two Alamouti codes, which lose their reduced complexity decoding on time-varying channels, we have shown that the golden code is fast decodable on both quasistatic and rapid time-varying channels.

APPENDIX A: PROOF OF THE KEY PROPERTY (THEOREM 1)

Recall from (6) that the effective channel matrix is $\mathbf{H} = \overline{\mathbf{H}}\Psi$. We will use a QR decomposition of $\overline{\mathbf{H}}$, namely $\overline{\mathbf{H}} = \overline{\mathbf{Q}}\,\overline{\mathbf{R}}$, to construct a QR decomposition of $\mathbf{H}$, namely $\mathbf{H} = \mathbf{Q}\mathbf{R}$.

Inspection of (6) reveals that $\overline{\mathbf{h}}_1^*\overline{\mathbf{h}}_2 = \overline{\mathbf{h}}_1^*\overline{\mathbf{h}}_4 = \overline{\mathbf{h}}_2^*\overline{\mathbf{h}}_3 = \overline{\mathbf{h}}_3^*\overline{\mathbf{h}}_4 = 0$, which implies that the subspace spanned by the first and third columns of $\overline{\mathbf{H}}$ is orthogonal to the subspace spanned by the second and fourth columns. This fact implies that $\overline{r}_{1,2} = \overline{r}_{1,4} = \overline{r}_{2,3} = \overline{r}_{3,4} = 0$, so that:

$$\mathbf{H} = \overline{\mathbf{H}}\Psi = \overline{\mathbf{Q}} \begin{bmatrix} \overline{r}_{11} & 0 & \overline{r}_{13} & 0 \\ 0 & \overline{r}_{22} & 0 & \overline{r}_{24} \\ 0 & 0 & \overline{r}_{33} & 0 \\ 0 & 0 & 0 & \overline{r}_{44} \end{bmatrix} \begin{bmatrix} c & s & 0 & 0 \\ -s & c & 0 & 0 \\ 0 & 0 & c & s \\ 0 & 0 & -s & c \end{bmatrix} \tag{22}$$

$$= \overline{\mathbf{Q}}\mathbf{G}, \tag{23}$$

where $\quad \mathbf{G} = \begin{bmatrix} \mathbf{X} & \mathbf{Y} \\ \mathbf{O} & \mathbf{Z} \end{bmatrix}, \mathbf{X} = \begin{bmatrix} c\overline{r}_{11} & s\overline{r}_{11} \\ -s\overline{r}_{22} & c\overline{r}_{22} \end{bmatrix}, \mathbf{Y} = \begin{bmatrix} c\overline{r}_{13} & s\overline{r}_{13} \\ -s\overline{r}_{24} & c\overline{r}_{24} \end{bmatrix}$, and $\mathbf{Z} = \begin{bmatrix} c\overline{r}_{33} & s\overline{r}_{33} \\ -s\overline{r}_{44} & c\overline{r}_{44} \end{bmatrix}. \tag{24}$

Observe that the submatrices $\mathbf{X}$ and $\mathbf{Z}$ are entirely real, since $c$ and $s$ and $\{\overline{r}_{ii}\}$ are all real. Therefore, we can transform $\mathbf{G}$ into an upper triangular matrix $\mathbf{R} = \mathbf{W}\mathbf{G}$ via the purely real Givens rotation matrix:

$$\mathbf{W} = \begin{bmatrix} \mathbf{W}_1 & \mathbf{O} \\ \mathbf{O} & \mathbf{W}_2 \end{bmatrix}, \tag{25}$$

where $\mathbf{W}_1 = \dfrac{1}{\sqrt{(c\overline{r}_{11})^2 + (s\overline{r}_{22})^2}} \begin{bmatrix} c\overline{r}_{11} & -s\overline{r}_{22} \\ s\overline{r}_{22} & c\overline{r}_{11} \end{bmatrix}$ and $\mathbf{W}_2 = \dfrac{1}{\sqrt{(c\overline{r}_{33})^2 + (s\overline{r}_{44})^2}} \begin{bmatrix} c\overline{r}_{33} & -s\overline{r}_{44} \\ s\overline{r}_{44} & c\overline{r}_{33} \end{bmatrix}$.

Substituting $\mathbf{G} = \mathbf{W}^T\mathbf{R}$ into (23) yields the desired QR decomposition $\mathbf{H} = \mathbf{Q}\mathbf{R}$, where $\mathbf{Q} = \overline{\mathbf{Q}}\mathbf{W}^T$ and

$$\mathbf{R} = \mathbf{W}\mathbf{G} = \begin{bmatrix} \mathbf{W}_1 & \mathbf{O} \\ \mathbf{O} & \mathbf{W}_2 \end{bmatrix} \begin{bmatrix} \mathbf{X} & \mathbf{Y} \\ \mathbf{O} & \mathbf{Z} \end{bmatrix} \tag{26}$$

$$= \begin{bmatrix} \mathbf{A} & \mathbf{B} \\ \mathbf{O} & \mathbf{D} \end{bmatrix}. \tag{27}$$

Since $\mathbf{W}_1$, $\mathbf{W}_2$, $\mathbf{X}$ and $\mathbf{Z}$ are all real, it follows that both $\mathbf{A} = \mathbf{W}_1\mathbf{X}$ and $\mathbf{D} = \mathbf{W}_2\mathbf{Z}$ are real. And by construction of $\mathbf{W}_1$ and $\mathbf{W}_2$, both $\mathbf{A}$ and $\mathbf{D}$ are upper triangular.

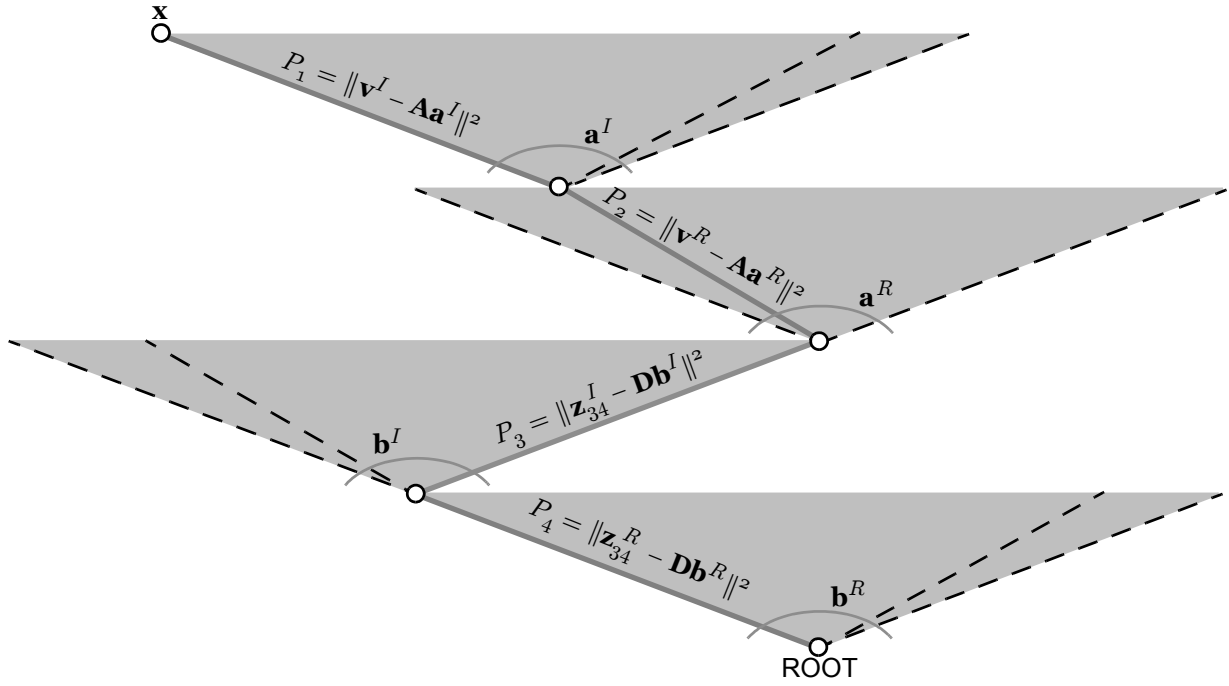

**Fig. 1.** The structure of the proposed detection tree and its branch metrics. The cost function for the leaf node is the sum of the branch metrics, $P(\mathbf{x}) = P_1 + P_2 + P_3 + P_4$.

1.          $[\mathbf{Q}, \mathbf{R}] = $ QR decomposition($\mathbf{H}$)
2.          $\mathbf{z} = \mathbf{Q}^*\mathbf{y}$
3.          $\hat{P} = \infty$
4.          $[P_4, \Pi_4] = $ **sort**$_{a \in \mathcal{A}}$( $(z_3^R - r_{33}a^R - r_{34}a^I)^2 + (z_4^R - r_{44}a^I)^2$ )
5.          $[P_3, \Pi_3] = $ **sort**$_{a \in \mathcal{A}}$( $(z_3^I - r_{33}a^R - r_{34}a^I)^2 + (z_4^I - r_{44}a^I)^2$ )
6.          **for** $k$ **from** 1 **to** $M$          % step through each $\mathbf{b}^R$ in order
7.              **if** $P_4(k) > \hat{P}$, **break**, **end**
8.              **for** $l$ **from** 1 **to** $M$     % step through each $\mathbf{b}^I$ in order
9.                  **if** $P_3(l) + P_4(k) > \hat{P}$, **break**, **end**
10.                $[x_3^R \; x_3^I \; x_4^R \; x_4^I] = [\mathcal{A}(\Pi_4(k))^R \mathcal{A}(\Pi_3(l))^R \mathcal{A}(\Pi_4(k))^I \mathcal{A}(\Pi_3(l))^I]$
11.                $v_1 = z_1 - r_{13}x_3 - r_{14}x_4$
12.                $v_2 = z_2 - r_{23}x_3 - r_{24}x_4$
13.                $\hat{P}_1 = \hat{P}_2 = \infty$
14.                $\mathcal{X} = $ **list**($v_2^R/r_{22}$)         % ordered list of decisions for $x_2^R$
15.                **for** $m$ **from** 1 **to** $\sqrt{M}$    % step through each $x_2^R$ in order
16.                    $P_2 = (v_2^R - r_{22}\mathcal{X}(m))^2$
17.                    **if** $P_2 > \hat{P}_2$, **break**, **end**
18.                    $u_1^R = v_1^R - r_{12}\mathcal{X}(m)$
19.                    $q = Q(u_1^R/r_{11})$
20.                    $P_2 = (u_1^R - r_{11}q)^2 + P_2$
21.                    **if** $P_2 < \hat{P}_2$, $x_1^R = q$, $x_2^R = \mathcal{X}(m)$, $\hat{P}_2 = P_2$, **end**
22.                **end**
23.                $\mathcal{X} = $ **list**($v_2^I/r_{22}$)         % ordered list of decisions for $x_2^I$
24.                **for** $n$ **from** 1 **to** $\sqrt{M}$    % step through each $x_2^I$ in order
25.                    $P_1 = (v_2^I - r_{22}\mathcal{X}(n))^2$
26.                    **if** $P_1 > \hat{P}_1$, **break**, **end**
27.                    $u_1^I = v_1^I - r_{12}\mathcal{X}(n)$
28.                    $q = Q(u_1^I/r_{11})$
29.                    $P_1 = (u_1^I - r_{11}q)^2 + P_1$
30.                    **if** $P_1 < \hat{P}_1$, $x_1^I = q$, $x_2^I = \mathcal{X}(n)$, $\hat{P}_1 = P_1$, **end**
31.                **end**
32.                $P = \hat{P}_1 + \hat{P}_2 + P_3(l) + P_4(k)$
33.                **if** $P < \hat{P}$, $\hat{\mathbf{x}} = [x_1, x_2, x_3, x_4]$, $\hat{P} = P$, **end**
34.              **end**
35.          **end**

**Fig. 2.** Pseudocode of a fast ML decoder for the golden code.

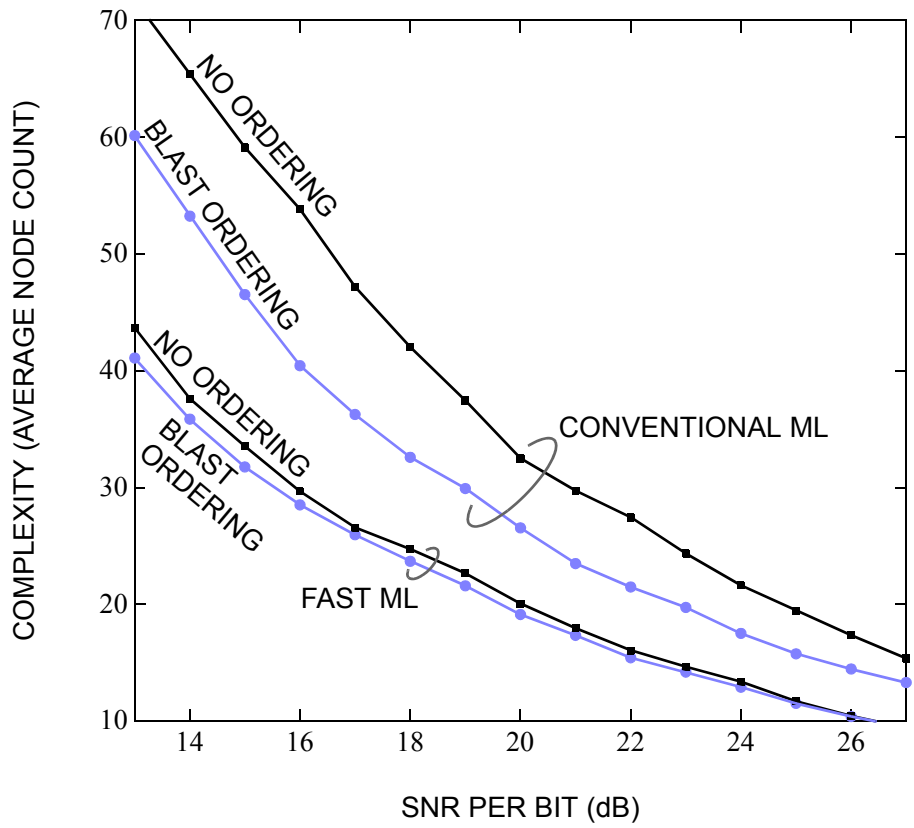

**Fig. 3.** Decoding complexity versus SNR for golden code with 64-QAM.